\documentclass[12pt]{iopart}
\usepackage{graphicx}
\usepackage{amssymb}
\usepackage{color}
\usepackage{cite}
\usepackage{hyperref}
\usepackage{subfigure}
\usepackage{epsfig}
\usepackage{xcolor}
\usepackage{bm}
\usepackage{latexsym}
\def\s{\sigma}
\def\om{\omega}

\def\da{\dagger}

\newcommand{\ket}[1]{|#1\rangle}
\newcommand{\abs}[1]{\left| #1 \right|} 
\newcommand{\expect}[1]{\mathinner{\langle #1\rangle}}

\def\beq{\begin{equation}}
\def\eeq{\end{equation}}
\def\beqa{\begin{eqnarray}}
\def\eeqa{\end{eqnarray}}
\def\bfig{\begin{figure}}
\def\efig{\end{figure}}

\newcommand{\eq}[1]{Eq.~(#1)}
\newcommand{\eqs}[1]{Eqs.~(#1)}
\newcommand{\fig}[1]{Fig.~#1}

\begin{document}
\title{Quantum criticality and state engineering in the simulated anisotropic quantum Rabi model}
\author{Yimin Wang$^{1,2}$, Wen-Long You$^{3}$, Maoxin Liu$^{1}$, Yu-Li Dong$^{3}$, Hong-Gang Luo$^{1,4}$, G. Romero$^{5}$, J. Q. You$^{1}$}

\address{$^1$ Quantum Physics and Quantum Information Division, Beijing Computational Science Research Center, Beijing 100094, China}
\address{$^2$ College of Communications Engineering, Army Engineering University, Nanjing 210007, China}
\address{$^3$ College of Physics, Optoelectronics and Energy, Soochow University, Suzhou, Jiangsu 215006, China}
\address{$^4$ Center for Interdisciplinary Studies $\&$ Key Laboratory for Magnetism and Magnetic Materials of the MoE, Lanzhou University, Lanzhou 730000, China}
\address{$^5$ Departamento de F\'{\i}sica, Universidad de Santiago de Chile (USACH), Avenida Ecuador 3493, 917-0124, Santiago, Chile}
\ead{wlyou@suda.edu.cn;jqyou@csrc.ac.cn}

\begin{abstract}
Promising applications of the anisotropic quantum Rabi model (AQRM) in broad parameter ranges are explored, which is realized with superconducting flux qubits simultaneously driven by two-tone time-dependent magnetic fields. Regarding the quantum phase transitions (QPTs), with assistant of fidelity susceptibility, we extract the scaling functions and the critical exponents, with which the universal scaling of the cumulant ratio is captured with rescaling of the parameters due to the anisotropy. Moreover, a fixed point of the cumulant ratio is predicted at the critical point of the AQRM. In respect to quantum information tasks, the generation of the macroscopic Schr\"{o}dinger cat states and quantum controlled phase gates are investigated in the degenerate case of the AQRM, whose performance is also investigated by numerical calculation with practical parameters. Therefore, our results pave a way to explore distinct features of the AQRM in circuit QED systems for QPTs, quantum simulations and quantum information processings. 
\end{abstract}


\maketitle
\section{Introduction}
Recent experimental progresses in solid-state-based quantum systems have allowed the advent of the so-called ultrastrong coupling (USC) regime \cite{niemczyk_circuit_2010,forn-diaz_observation_2010,chen_single-photon-driven_2017} and the deep strong coupling (DSC) regime \cite{yoshihara_superconducting_2017,forn-diaz_ultrastrong_2017} of light-matter interactions, where the coupling strength is comparable to (USC) or larger than (DSC) appreciable fractions of the mode frequency. In these regimes, the celebrated rotating-wave approximation (RWA) breaks down and the quantum Rabi model (QRM) is invoked\cite{rabi_process_1936,braak_integrability_2011}.
In addition to the relatively complex quantum dynamics provided by the QRM, it brings about novel quantum phenomena \cite{ridolfo_photon_2012,sanchez-burillo_scattering_2014,hwang_quantum_2015} and challenges in implementing quantum information tasks \cite{romero_ultrafast_2012,kyaw_scalable_2015,wang_holonomic_2016,wang_ultrafast_2017}. Although exciting, natural implementations of the QRM in the USC/DSC regime in other platforms remain very challenging since they are confined by fundamental limitations. However, different schemes have been used to simulate the QRM using superconducting circuits~\cite{ballester_quantum_2012,langford2017nc}, quantum optical systems~\cite{crespi_photonic_2012}, trapped ions~\cite{pedernales_quantum_2015,lv2017app} and cold atoms~\cite{felicetti_quantum_2017}.

In the other aspect, the fascinating promises of the QRM has trigged many studies of the anisotropic quantum Rabi model (AQRM), see e.g., Refs. \cite{xie_anisotropic_2014,zhang2017praa,liu_universal_2017},
\beq
H_{AQRM} = \hbar \tilde{\om} a^\da a +  \frac{\hbar}{2}\tilde{\om}_q \s_z +\hbar \tilde{g}[ (\s_- a^\da+\s_+ a  )+\hbar \tilde{\lambda} (\s_+ a^\da + \s_- a)] ,
\label{eq_harabi}
\eeq
where $a$ and $a^\dag$ are the annihilation and creation operators of the bosonic mode with frequency $\tilde{\om}$, $\s_z$ and $\s_x=\s_+ + \s_-$ are Pauli operators associated with a qubit with ground state $\ket{g}$, excited state $\ket{e}$, and transition frequency $\tilde{\om}_q$. It is a generalization of the QRM affiliated with the $\tilde{\lambda}$ denoting the asymmetry between rotating and counter-rotating terms. The AQRM returns to the Jaynes-Cummings model (JCM) with $\tilde{\lambda}=0$~\cite{jaynes_comparison_1963}, or the original QRM with $\tilde{\lambda} =1$.

Since individual addressing of the two coupling constants are allowed, the AQRM presents a favorable test bed for many valuable theoretical issues, such as the role of the counter-rotating terms~\cite{huang_phase-kicked_2015,zhang_analytical_2015,wu_effects_2017}, the Fisher information \cite{wang_energy-level_2016}, the universality scaling of the quantum phase transition (QPT), the enhanced squeezing~\cite{zhang2017praa}, and thus it may bridge the gap between the JCM and the QRM in the dynamics \cite{wang_bridging_2015}. The discussions of the anisotropy in the standard AQRM were further extended to the semi-classical case~\cite{dai2017apl}, the multi-qubit case~\cite{baksic_controlling_2014,liu_universal_2017} (namely the anisotropic Dicke model). As a consequence, the theoretical advancements bring up experimental requests for individual adjustability of the coupling constants $\tilde{g}_r$ and $\tilde{g}_{cr}$ in wide parameter ranges to demonstrate the innovative features of the AQRM. Although there have already been some experimental proposals for the realization of the AQRM in some systems, i.e., quantum well with spin-orbit coupling \cite{schliemann_variational_2003,wang_energy-level_2016}, and circuit QED systems \cite{baksic_controlling_2014,yang_ultrastrong-coupling_2017}, they are quite limited on the tunability and the achievable parameter ranges. Therefore, the demand to explore new platforms to study the dynamics of the AQRM is put forward.

In this work, we propose an experimentally feasible scheme to simulate the controllable AQRM demonstrating the USC and DSC dynamics with superconducting flux qubits. 
We show through analytical and numerical calculations that our schematic setup has the distinct advantage that the parameters in the effective AQRM can be individually controlled by the frequencies and the amplitudes of the bichromatic magnetic fluxes.  The all-round tunability of parameters in this model provides a powerful tool for exploring a few appealing issues. It should be aware that the present AQRM can not only reduce to the QRM but also produce the primitive JCM with a sufficiently strong coupling strength, which gives the opportunity of experimental observation of the gapless Nambu-Goldstone mode. Firstly we focus on the long-sought QPT in a few-body system, which is initially thought as a privilege of  quantum many-body systems, the critical phenomena and the universal properties of the effective AQRM can be addressed. The critical exponents can be extracted from the scaling behavior of the fidelity susceptibility. Independent of the diversities in the anisotropy and the frequency size, a fixed point of a cumulant ratio is predicted and a universal scaling of the cumulant ratio is obtained. Besides, two-qubit quantum gates and Schr\"{o}dinger cat states can be produced in the special degenerate case of the AQRM. Therefore, our proposal not only pave a way to implement quantum simulators \cite{georgescu_quantum_2014} and quantum information tasks, but also the way to explore the QPTs for rich coupling regimes of light-matter interaction in systems where they are experimentally inaccessible.

The paper is organized as follows. We firstly describe in Sec. \ref{sec2} the Hamiltonian of our qubit-resonator setup, where the flux qubit is controlled by bichromatic time-dependent magnetic fluxes. The effective AQRM is obtained when the frequency conditions are well-respected. In Sec. \ref{sec3}, we study the QPTs of the simulated AQRM with the method of fidelity susceptibility and scaling theory. In Sec. \ref{sec4}, we discuss quantum information applications with the degenerate AQRM, such as the generation of the macroscopic Schr\"{o}dinger cat states and the quantum controlled phase gates. The conclusions are presented in Sec. \ref{sec5}.

\section{The qubit-resonator circuit}
\label{sec2}
\begin{figure}[b]
\begin{center}
\includegraphics[scale=0.8]{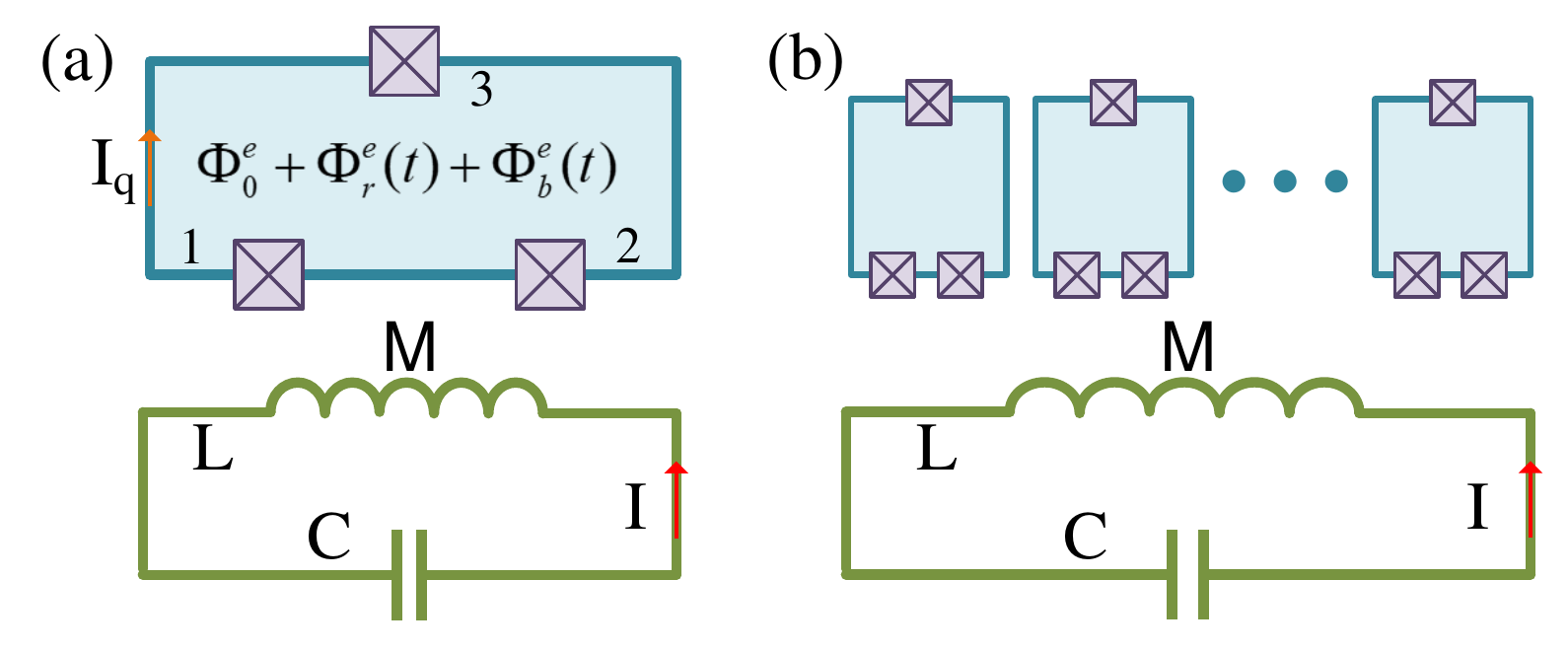}
\end{center}
\caption{Schematic representation of our setup. (a) A flux qubit with three Josephson junctions is coupled to a LC circuit by the mutual inductance $M$. The externally applied magnetic flux threading the qubit's loop, which includes a dc term $\Phi_0^{e}$ and two ac terms $\Phi_r^{e}(t)$, $\Phi_b^{e}(t)$, controls the qubit-resonator coupling. The currents through the qubit and the LC circuit are denoted by $I_q$ and $I$, respectively. (b) Scale up to the multi-qubit case.}	
\label{fig_setup}
\end{figure}
For simplicity, but here without loss of generality, we use three-junction flux qubits (e.g.,~\cite{orlando_superconducting_1999,liu_superconducting_2007}) in our scheme. As shown in \fig{\ref{fig_setup}}(a), a flux qubit is coupled to a LC circuit with an inductance $L$ and a capacitance $C$. The mutual inductance between the flux qubit and the LC circuit is $M$. The applied magnetic flux $\Phi$ through the flux qubit loop in \fig{\ref{fig_setup}}(a), which controls the qubit-resonator couplings, is assumed to include a static magnetic flux $\Phi_0^{\rm e}$, and also two time-dependent magnetic fields (TDMFs),~$\Phi_j^{\rm e}(t)=A_j\cos(\omega_j t+\varphi_j)$. Here $j=r,b$ label the two TDMFs, individually.
Considering one three-junction flux qubit, the qubit's Hamiltonian $H_q$ reads $H_q=\sum_{i=1}^{3}[C_{{\rm
J}i} \Phi^2_0\dot{\phi}_{i}^2/{(8\pi^{2})}-E_{{\rm J}i}\cos\phi_{i}]$, where we have assumed each junction in the flux qubit has a capacitance $C_{{\rm J}i}$, phase drop $\phi_{i}=2\pi\Phi_i/\Phi_0$, Josephson energy $E_{Ji}$, and critical current $I_{0i}=2\pi E_{{\rm J}i}/\Phi_{0}$. Here, $\Phi_{0}=h/2e$ is the magnetic flux quantum. With current-phase relation, the super-current for each junction reads $I_{i}=I_{0i}\sin\phi_{i}$. And thus the persistent current in the qubit loop is $I_q=C_q\sum_{i=1}^{3}I_{0i} \sin\phi_{i}/C_{{\rm J}i}$~\cite{liu_superconducting_2007,huang_optimal_2014}, where $C_q$ is the total capacitance of the flux qubit, $C_q^{-1}=\sum_i C_{{\rm J}i}^{-1}$, with the convention $C_{{\rm J}3}=\eta C_{{\rm J}1}=\eta C_{{\rm J}2}$ and $\eta$ being the relative size of the Josephson junction.
Taking into account the TDMFs, the flux quantization around the qubit's loop imposes a constraint on the phase drop across the three
junctions~\cite{orlando_superconducting_1999,liu_superconducting_2007,huang_optimal_2014},
$\sum_{i=1}^{3}\phi_{i}+2\pi(\Phi_0^e+\Phi_r^e(t)+\Phi_d^e(t))/{\Phi_0} =0$. In order to define an effective qubit within the junction architecture, we diagonalize the Hamiltonian $H_q$ containing only the junctions in absence of the TDMFs, i.e., $\Phi_j^e(t)=0$. The two lowest eigenstates are labeled as the eigenstates of $\sigma_z$, i.e., $\ket{g}$ and $\ket{e}$, and the spanned two-dimensional subspace describes the effective qubit.

In the other aspect, the Hamiltonian of the total system is written as $H=H_q+H_c+ I M I_q$, where $H_c={Q^2/2C}+{\Phi^2}/{2L}$ is the Hamiltonian of the LC circuit, with $Q$ being the capacitor's charge, $\Phi=IL$ being the magnetic flux through the LC circuit loop and $I$ is the inductor's current. The Hamiltonian of the LC circuit $H_c$ can be simply quantized by introducing the annihilation and creation operators
$\Phi=\sqrt{\hbar/(2 C \omega)} (a+a^\dag)$ and $Q=-i\sqrt{(\hbar C  \omega) /2} (a-a^\dag)$,
with the frequency $\omega=1/\sqrt{LC}$.
After projecting the total Hamiltonian into the qubit's bases $\{\ket{g}, \; \ket{e}\}$, we obtain
\beq
\tilde{H}=\frac{\hbar}{2}\omega_{q}\sigma_{z}+\hbar\omega a{^\dag}a +  \hbar g \sigma_x (a{^\dag}+a) +\sum_{j=r,b}\cos(\omega_j t + \varphi_j)  \left[\Omega_j\sigma_x -\Lambda_j \sigma_x (a^{\dagger}+a)\right].
\label{eq_ht}
\eeq
The first two terms in \eq{\ref{eq_ht}} denote the free Hamiltonians of both the qubit and the LC circuit, where $\omega_{q}$ is the transition frequency of the effective qubit. The third term in \eq{\ref{eq_ht}} represents the qubit-resonator interaction with the coupling strength being $g=M\sqrt{{\omega}/{2\hbar L}}|{\langle e|\tilde{I}_q|g\rangle}|$. Here $\tilde{I}_q=C_q\sum_{i=1}^{2}{I_{0,i}}\sin\phi_{i}/{C_{{\rm J} i}}+C_q{I_{0,3}}\sin{\tilde{\phi}_{3}}/{C_{{\rm J} 3}}$ is the super-current through the qubit loop when $\Phi_j^e(t)=0$, where $\tilde{\phi}_3 = -\left(\phi_r+\phi_b+2\pi f_e \right)$ and $f_e \equiv \Phi_0^e/\Phi_0$ is the reduced dc bias magnetic flux.
The fourth term in \eq{\ref{eq_ht}} plays the role of a driving Hamiltonian representing the interaction between the qubit and TDMFs with the respective driving strength being $\Omega_j =A_j \abs {\langle e|I_{03}\sin{\tilde{\phi}_3}|g\rangle}$
.
The fifth term of \eq{\ref{eq_ht}} is the controllable nonlinear interaction among the qubit, the resonator, and the TDMFs, with the respective coupling strength being $\Lambda_j={4\pi^2 A_j MC_q}{C^{-1}_{{\rm J}3}\Phi^{-2}_{0}}\sqrt{{\hbar\omega}/{2L}}\,|{\left\langle e\right|E_{{\rm J}3}\cos\tilde{\phi}_{3}\;\left|g\right\rangle}|$.
As noticed above, the TDMFs $\Phi^{e}_j(t)$ equal to zero when calculating the coupling strengths $g$, $\Omega_j$, and $\Lambda_j$.
It is worthy noting that in the above derivations, we keep the time-dependent amplitudes small such that the reduced time-dependent magnetic fluxes satisfy $\abs{f_j(t)}\lesssim 10^{-3}$. This leads to: (1) the approximation of $\sin[2\pi f_j(t)]\sim 2\pi f_j(t)$ and $\cos[2\pi \Phi_j^e(t)/\Phi_0]\sim 1$; (2) the ignorance of the interaction terms controlled by two simultaneously applied TDMFs in the form of $\sim f_1(t)f_2(t)$. As a result, when expanding the potential energy in qubit's Hamiltonian $H_q$ and the qubit's loop current $I_q$, we only need to keep the first order of the small reduced flux $\Phi^{e}_j(t)/\Phi_0$. 

With realistic parameters discussed in \cite{liu_superconducting_2007}, the frequency of the LC oscillator can be designed to be $\omega \approx 2\pi \times 3$~GHz, the qubit's frequency is approximately $\omega_q \approx 2\pi \times 18$~GHz, with $g \approx 2\pi \times 37$~MHz when $f_e = 0.49$. Therefore, the conditions $g\ll\abs{\omega_q \pm\omega} $ is well satisfied and the effect of the always-on qubit-resonator interaction term is negligibly small. \eq{\ref{eq_ht}} can be written as
\beqa
\label{eq_htotaltem}
\tilde{H}_{int}'=& \sum_{j=r,b} \Omega_j\cos(\omega_j t + \varphi_j) \left(\sigma_+ e^{i \omega_q  t} + h.c.\right) \nonumber \\
&-\sum_{j=r,b} \Lambda_j \cos{(\omega_j t + \varphi_j)} \left(a^{\dagger}e^{i \omega  t}+a e^{-i \omega  t}\right) \left(\sigma_+ e^{i \omega_q  t} + h.c.\right),
\eeqa
where we have performed the RWAs and neglected all terms that are fast oscillating in the interaction picture with respect to the system's free Hamiltonian $\hbar \omega_{q}\sigma_{z}/2+\hbar\omega a{^\dag}a$.
We next consider the case where the TDMFs are inducing the respective first-order red (r) and blue (b) sideband transitions with small detunings $\delta_r$ and $\delta_b$ onto the qubit-resonator system, e.g., $\omega_r = \omega_q - \omega - \delta_r$ and $\omega_b =  \omega_q + \omega - \delta_b$. In such a scenario, the terms in the first line of \eq{\ref{eq_htotaltem}} representing the direct driving on the qubit can be ignored for weak drivings such that $\Omega_j/\hbar\ll {\rm min}\{\omega_q,\abs{\omega_q \pm\omega_j}\}$ is fulfilled, since $\Omega_j/\hbar$ is a few megahertz and $\abs{\omega_q \pm\omega_j}$ is on the order of gigahertz. Similarly, when the rest of the frequency detunings are large compared to the coupling parameters, i.e.,  $\abs{\omega - \omega_r \pm \omega_q}\gg\Lambda_r$ and $ \abs{\omega -\omega_q \pm \omega_b}\gg\Lambda_b $, one may neglect the rest of the fast-oscillating terms. These approximations lead to a simplified time-dependent Hamiltonian
\beq
\tilde{H}_{eff}=\frac{\Lambda_r}{2}\left(\sigma_+ a\, e^{i \delta_r t + i\varphi_r}+h.c.\right)+\frac{\Lambda_b}{2} \left(\sigma_+ a^\dag\, e^{i \delta_b t+i\varphi_b}+h.c.\right),
\label{eq_anrabiip}
\eeq
It is worth noting that \eq{\ref{eq_anrabiip}} corresponds to the interaction picture of the generalized AQRM with respect to the uncoupled Hamiltonian $H_0=\hbar\left(\delta_r+\delta_b\right)\sigma_z /4+\hbar \left(\delta_b-\delta_r\right)a^\dag a/2$~ \cite{pedernales_quantum_2015}, such that
\beq
H'_{AQRM}=\frac{\hbar}{2} \tilde{\omega}_q\sigma_z + \hbar \tilde{\omega} a^\dag a +\hbar \tilde{g}_r\Big(\sigma_+ a \,e^{ i\varphi_r} + h.c.\Big) +\hbar \tilde{g}_{cr}\Big(\sigma_+ a^\dag \, e^{i\varphi_b} + h.c.\Big),
\label{eq_heff}
\eeq
with the effective parameters being $\tilde{\omega}_q= (\delta_r+\delta_b)/2, \tilde{\omega}=(\delta_b-\delta_r)/2,
\tilde{g}_r = {\Lambda_r}/{(2\hbar)},\tilde{g}_{cr} = {\Lambda_b}/{(2\hbar)}.$
Here the qubit's and the resonator's frequencies are represented by the sum and the difference of the two detunings, respectively. The tunability of these parameters permits the study of all coupling regimes of the AQRM via suitable choices of the amplitudes and the detunings of the TDMFs. It is noteworthy that the complex coupling strengths $\tilde{g}_r$ and $\tilde{g}_{ar}$ can be realized by choosing the phases $\varphi_r$ and $\varphi_b$ of the TDMFs. For example, \eq{\ref{eq_heff}} leads to the standard AQRM in \eq{\ref{eq_harabi}} when $\varphi_r=\varphi_b=0$ and $\tilde{\lambda}=\tilde{g}_{cr}/\tilde{g}_{r}$. And to go beyond the USC, we only need the condition that $\Lambda_j\gtrsim \hbar \abs{\delta_b-\delta_r}$. In particular, with only one single frequency TDMF on the flux qubit, i.e., $\Lambda_b = 0$, \eq{\ref{eq_heff}} reduces to standard JCM, which possesses the continuous $U(1)$ symmetry. One interesting point is that it provides different phase diagrams with sufficiently strong coupling strength, such as the emergence of the gapless excitation spectrum, namely the so-called Nambu-Goldstone mode~\cite{fan2014pra,hwang2016prl}. It should be emphasized that a pure JCM with a very large coupling strength does not naturally exist due to the breakdown of the RWA, which in a way prevents the experimental observation of the Nambu-Goldstone modes. Therefore, our scheme sheds bright light on the possibility of experimentally demonstrating the gapless excitation spectrum in circuit QED systems.

\section{Quantum phase transition and finite frequency scaling in AQRM}
\label{sec3}
Quantum phase transition in systems with few degrees of freedom has been a topic of major interest recently~\cite{hwang_quantum_2015,liu_universal_2017,wei2018prab}. Although the quantum criticality is commonly believed to take place in a many-body system in the thermodynamic limit, 
it is newly realized that a few-body system may undergo a QPT provided the energy barrier between two local-minima states is infinite~\cite{hwang_quantum_2015,liu_universal_2017,wei2018prab}. 
Since the simulated AQRM \eq{\ref{eq_heff}} in hand, the free adjustment of parameters allows us to investigate the interesting issues like the critical phenomena and the universal properties. In this section, we study the QPTs in the AQRM from the perspective of fidelity susceptibility and the scaling theory.

\subsection{Fidelity susceptibility with AQRM}
The fidelity susceptibility \cite{you2007pre} is originally proposed to elucidate the changing rate of fidelity \cite{quan2006prl,zanardi2006pre} under an infinitesimal variation of the driving parameter:
\begin{eqnarray}
\label{eq7}
|\langle\Psi_G(\tilde{g})|\Psi_G(\tilde{g}+\delta\tilde{g})\rangle|  =1-\chi_F \delta \tilde{g}^2/2 + O(\delta \tilde{g}^3),\label{eq_fidedifintion}
\end{eqnarray}
where $\vert \Psi_G(\tilde{g}) \rangle$ is the ground-state wave function of a Hamiltonian $H(\tilde{g})$=$H_0$ + $\tilde{g} H_I$, $\tilde{g}$ is the external driving parameter, and $\delta \tilde{g}$ is a tiny variation of the external parameter. Though borrowed from the quantum information theory, the fidelity susceptibility has been proved to be an effective sensor to detect and characterize QPTs in condensed matter physics \cite{gu2010ijmpb}. As an informational metric, the quantum fidelity susceptibility can be also devised to seize the criticality in the perspective of the Riemannian metric tensor form
\begin{eqnarray}
\chi_{\rm F}=\langle \partial_{\tilde{g}} \Psi_G \vert \partial_{\tilde{g}} \Psi_G \rangle - \vert  \langle \partial_{\tilde{g}} \Psi_G \vert \Psi_G \rangle \vert^2.
\label{eq_chipartialform}
\end{eqnarray}
There is a refreshing proposed duality, which  connects the fidelity susceptibility and the max volume of a codimension-one time slice in anti–de Sitter (AdS) space~\cite{miyaji2015prl,gan2017prd}. Such duality bridges quantum information theory and holography, and may deepen our understanding of quantum gravity \cite{susskind2016fpa,susskind2016fp}.


\begin{figure}[!b]
\begin{center}
\includegraphics[scale=0.8]{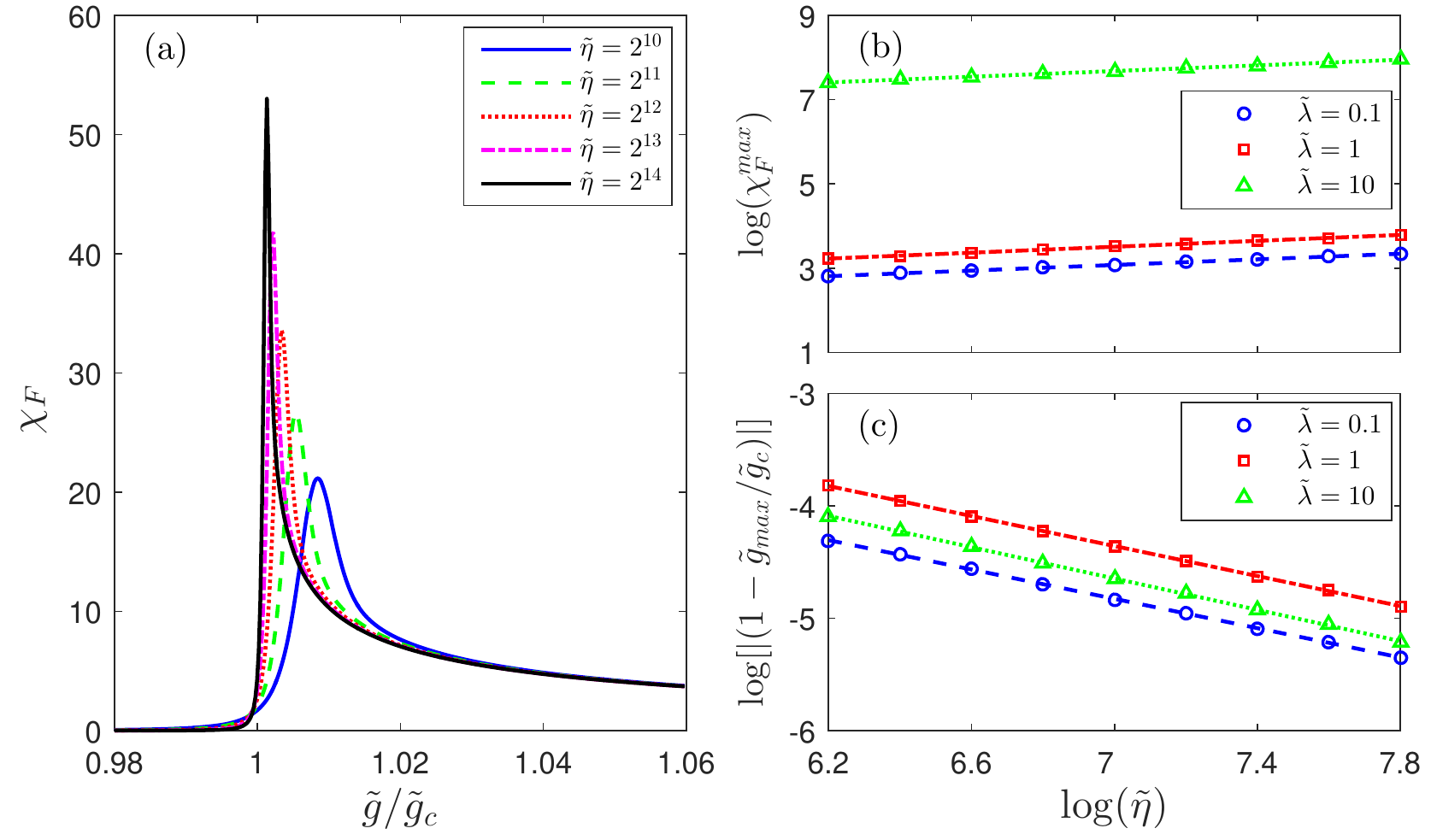}
\end{center}
\caption{(a)~Fidelity susceptibility in the AQRM as a function of the normalized coupling strength $\tilde{g}/\tilde{g}_c$ for different values of $\tilde{\eta}$ with $\tilde{\lambda} = 0.1$. (b)~The finite frequency-$\tilde{\eta}$ of the maximized fidelity susceptibility, $\log(\chi_F^{max})$ as a function of $\log(\tilde{\eta})$. With the linear fits of the numerical data, we find $\chi_F^{max}\approx\tilde{\eta}^{\,d_a^c}$ with $d_a^c\approx0.333$ for $\tilde{\lambda} = 0.1$ (blue dashed line), $d_a^c\approx0.351$ for $\tilde{\lambda} = 1$ (red dashed-dotted line), $d_a^c\approx0.337$ for $\tilde{\lambda} = 10$ (green dotted line). (c) The position of the maximized fidelity susceptibility, $\log[\abs{1-\tilde{g}_{max}/\tilde{g}_c}]$ as a function of $\log(\tilde{\eta})$. The linear fittings to the numerical data leads to $\abs{1-\tilde{g}/\tilde{g}_c)}\sim\tilde{\eta}^{-1/\nu}$ with $1/\nu\approx0.652$ for $\tilde{\lambda} = 0.1$ (blue dashed line), $1/\nu\approx0.668$ for $\tilde{\lambda} = 1$ (red dashed-dotted line), $1/\nu\approx0.697$ for $\tilde{\lambda} = 10$ (green dotted line).}
\label{fig_fs}
\end{figure}

To facilitate the QPTs of the AQRM in \eq{\ref{eq_harabi}}, we introduce a frequency ratio $\tilde{\eta}=\tilde{\omega}_q/\tilde{\omega}_r$  and the modified critical coupling strength $\tilde{g}_c=\sqrt{\tilde{\omega}_q \tilde{\omega}_r}/(1+\tilde{\lambda})$.  For a second-order QPT, around the critical point $\tilde{g}_c$, the correlation length $\xi$ diverges as $(\tilde{g} -\tilde{g}_c)^{-\nu}$, while the gap in the excitation spectrum vanishes as $(\tilde{g}-\tilde{g}_c)^{z\nu}$, where $\tilde{g}_c$ is the critical point, $\nu$ and $z$ are the correlation-length and dynamic exponents, respectively. We can account for the divergence around quantum critical points (QCPs) in the AQRM by formulating a finite-frequency scaling theory, in parallel with the scaling theory for finite-size effects in a many-body system, and here $\tilde{\eta}$ plays a similar role as system size in the latter case. Universal information could be decoded from the scaling behavior of the fidelity susceptibility \cite{kwok2008prea,gu2008prba,yu2009prea}. The fidelity susceptibility exhibits stronger dependence on $\tilde{\eta}$ across the critical point than in the non-critical region. Referring to standard arguments in finite-size scaling analysis \cite{continentino2001}, one obtains that the fidelity susceptibility can exhibit a finite-$\tilde{\eta}$ scaling. For finite-$\tilde{\eta}$ system, the position of a divergence peak defines a pseudocritical point $\tilde{g}_{max}$ as the precursor of a QPT. And it approaches the critical point $\tilde{g}_c$ as $\tilde{\eta} \to \infty$, which implies an abrupt change in the ground state of the system at the QCP in the in the classical oscillator limit. The maximum point of the fidelity susceptibility at $\tilde{g}_{max}$ scales like
\begin{eqnarray}
\chi_{\textrm{F}}(\tilde{g}_{max}) \sim \tilde{\eta}^{\,d_a^c},
\label{eq_chiscaling}
\end{eqnarray}
where $d_a^c$ denotes the critical adiabatic dimension. On the other hand, the position of the pseudocritical point obeys such scaling behavior as
\begin{eqnarray}
|\tilde{g}_{max} - \tilde{g}_c | \sim \tilde{\eta}^{-1/\nu}.
\label{eq_gscaling}
\end{eqnarray}
Thus, the behavior of $\chi_{\textrm{F}}$ on finite systems in the vicinity of a second-order QCP can be estimated as
\begin{eqnarray}
\chi_{\textrm{F}}  \approx C \tilde{\eta}^{d_a^c} \,f[\abs{\tilde{g}-\tilde{g_c}}\,\tilde{\eta}^{1/\nu}],
\label{eq_fscollapsescaling}
\end{eqnarray}
where $f$ is an unknown regular scaling function, and $C$ is a constant independent of $\tilde{g}$ and $\tilde{\eta}$.

In \fig{\ref{fig_fs}}~(a), we show the fidelity susceptibility $\chi_F$ in the AQRM as a function of the normalized coupling strength $\tilde{g}/\tilde{g}_c$ for different values of $\tilde{\eta}$ with $\tilde{\lambda} = 0.1$. Relatively away from critical points, the fidelity susceptibility $\chi_{\textrm{F}}(\tilde{g})$ is independent of $\tilde{\eta}$. The fidelity susceptibilities are always peaked around the critical point of AQRM, i.e.,~$\tilde{g}_{max}=\tilde{g}_c$. As $\tilde{\eta}$ increases, the values of the fidelity susceptibility becomes larger and the distribution becomes narrower, and the coupling strength $\tilde{g}_{max}$ corresponding to the maximized fidelity susceptibility approaches to the quantum critical point $\tilde{g}_c$ of the AQRM, which separate a normal phase at $\tilde{g} \le \tilde{g}_c$ from a super-radiant phase at $\tilde{g} \ge \tilde{g}_c$.

To extract the critical exponents $d_a^c$ and $\nu$ appearing in \eq{\ref{eq_chiscaling}} and \eq{\ref{eq_gscaling}}, we next study the finite-frequency scalings of the fidelity susceptibility for different values of $\tilde{\lambda}=0.1,\,\,\,1,\,\,\,10$, as shown in \fig{\ref{fig_fs}}(b) and \fig{\ref{fig_fs}}~(c). Here in both figures, the colored-markers  represent the calculated numerical results, and the colored-lines are the numerical fittings to the corresponding data sets.
The maximal fidelity susceptibility $\chi_F^{max}$ follows a universal power law with respect to $\tilde{\eta}$, as indicated from \fig{\ref{fig_fs}}(b). The slopes of the fitting indicate that the adiabatic dimension $d_a^c$ is $1/3$ ~\cite{hwang_quantum_2015,liu_universal_2017}.
Figure \ref{fig_fs}(c) exhibits the position of the coupling strength corresponding to $\log(\chi_F^{{max}})$, i.e., $\tilde{g}_{max}$, as a function of $\log{(\tilde{\eta})}$.
Similarly, the slopes of the fitting lines in suggest the value of the other critical exponent $\nu\approx3/2$. 
The gap scaling shows that $z\nu=1/2$ 
\cite{hwang_quantum_2015,liu_universal_2017}, and this accounts for that the dynamic exponent $z=1/3$, which agrees well with the excitation energy according to $\epsilon \sim \eta^{-z}$.

\subsection{The cumulant ratio and the fixed point with AQRM}
\begin{figure}[!b]
\begin{center}
\includegraphics[scale=0.8]{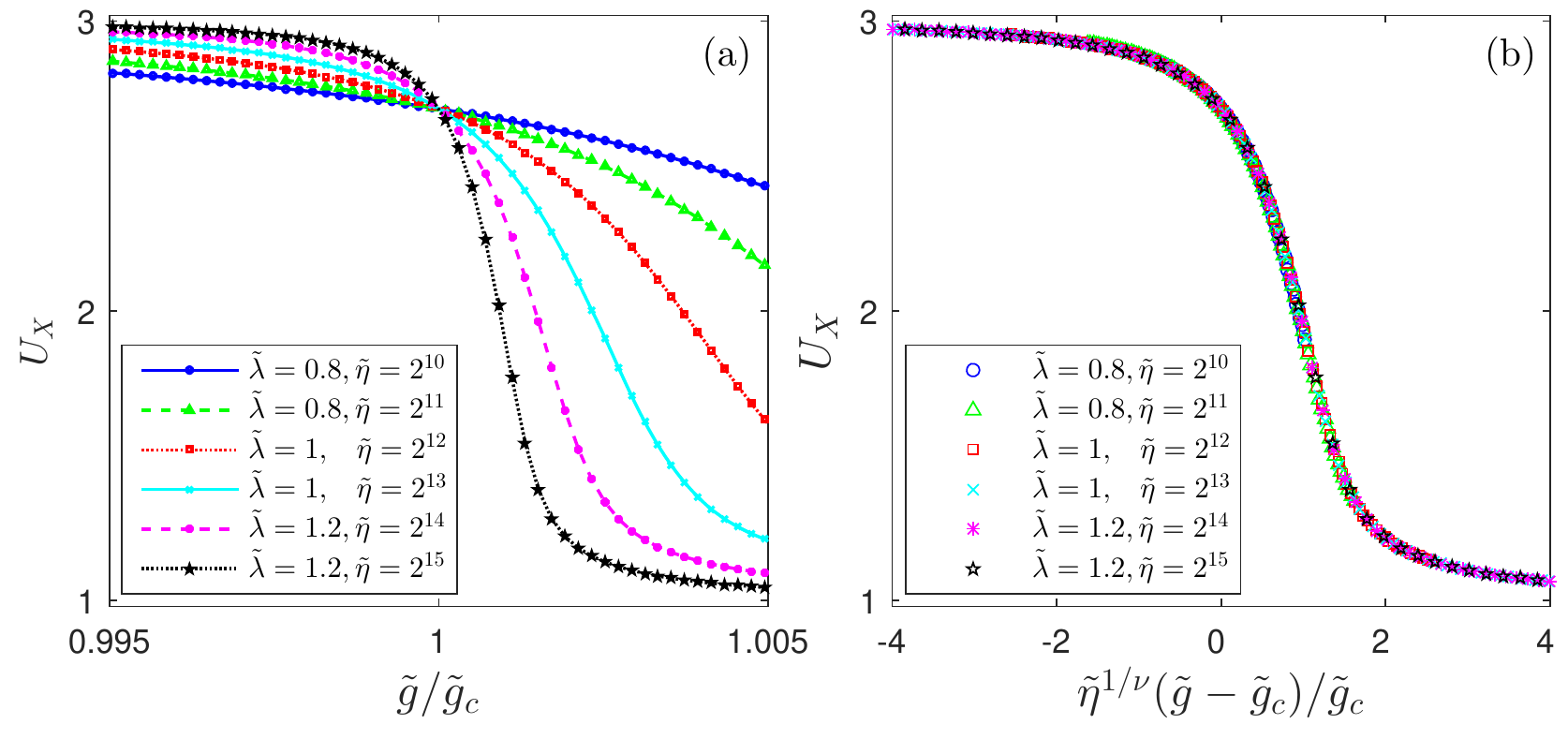}
\end{center}
\caption{(a)~The cumulant ratio $U_X$ as a function of the normalized coupling strength $\tilde{g}/\tilde{g}_c$ for different values of $\tilde{\eta}$ and $\tilde{\lambda}$. All the lines coincide with each other at the fixed point of $\tilde{g}/\tilde{g}_c$. (b)~The universal scaling of the cumulant ratio $U_X$ as a function of $\tilde{\eta}^{1/\nu}(\tilde{g}-\tilde{g}_c)/\tilde{g}_c$ for different values of $\tilde{\eta}$ and $\tilde{\lambda}$. All numerical data collapse into the single function.}
\label{fig_uxsp}
\end{figure}
With the extracted values of the critical exponents, i.e., $d_a^c=1/3$ and $\nu=3/2$, we are now ready to discuss the finite-frequency scaling properties of the field's position quadrature operator, $X_r = (a + a^\dagger)/\sqrt{2}$.
Taking average over the ground state of the AQRM, the scaling behavior of the field's position quadrature can be written in the form of~\cite{liu_universal_2017}
\beq
\expect{X_r^{2n}} = \tilde{\eta}'^{d_a^c}\, {\mathcal X}_n[(\tilde{g}-\tilde{g}_c)\,\tilde{\eta}'^{1/\nu}],
\label{eq_xrscaling}
\eeq
Here, $\tilde{\eta}' = \tilde{\eta}\,{(1+\tilde{\lambda})}/{(2\sqrt{|\tilde{\lambda}}|)}$ is the relative scaling variable modified by the anisotropy parameter $\tilde{\lambda}$, and ${\mathcal X}_n$ is the universal scaling function.
The scaling form of Eq. \eq{\ref{eq_xrscaling}} includes the anisotropic effect, which is beyond the traditional scaling frame. Actually, this anisotropic involved universality is urgently needed the experimental verification. For this purpose, we design a cumulant ratio as
\beq
U_X= \frac{\expect{X_r^{4}}}{\expect{X_r^{2}}^2}
= \frac{{\mathcal X}_2 [({\tilde{g}}-{\tilde{g}_c})\tilde{\eta}'^{1/\nu}]}{{\mathcal X}_1 ^2[(\tilde{g}-{\tilde{g}_c})\tilde{\eta}'^{1/\nu}]},
\label{eq_uxscaling}
\eeq
which can be considered as an analogy to the famous Binder cumulant ratio of the dimensional criticality in a statistical system~\cite{binder1981prl,binder1981zpb-cm}. Obviously, at the critical point $\tilde{g}=\tilde{g}_c$, $U_X$ is independent of $\tilde{\eta}'$, and locates at a universal value of  $\mathcal{X}_2(0)/\mathcal{X}_1^2(0)$. It means there is a fixed point at $\{\tilde{g}_c, \mathcal{X}_2(0)/\mathcal{X}_1^2(0)\}$, which is universal for different values of anisotropic strength $\tilde{\lambda}$ and frequency ratio scale $\tilde{\eta}'$. We confirm this fixed point involved universality via numerical calculation.
In fact,  we show in \fig{\ref{fig_uxsp}}~(a) that, the numerical values of $U_X$ for different values of $\tilde{\lambda}$ and $\tilde{\eta}'$ crossover each other at the fixed point of $\tilde{g}/\tilde{g}_c=1$, which is exactly the critical point of the QPT in the AQRM. Moreover, \fig{\ref{fig_uxsp}}~(b) implies that, despite the differences in $\tilde{\lambda}$ and $\tilde{\eta}'$, the cumulant ratio $U_X$ collapse into a single curve when appropriately scaled. This unambiguously reveals the observable-dependent scaling function ${\mathcal X}_n$ in \eq{\ref{eq_uxscaling}}.
Therefore, we can draw the same conclusion as that in Ref. \cite{liu_universal_2017}, which is, in a word, the QPTs in the AQRM and QRM are in the same universality class. However, the defined cumulant ratio $U_X$, which only needs to measure the quadrature of the displacement, has the advantage of experimental convenience. Moreover, we hope that the fixed point discussion presents an alternative possibility to explore the universality of the AQRM.

\subsection{The simulated AQRM with finite large frequency}
\begin{figure}[!b]
\begin{center}
\includegraphics[scale=1.2]{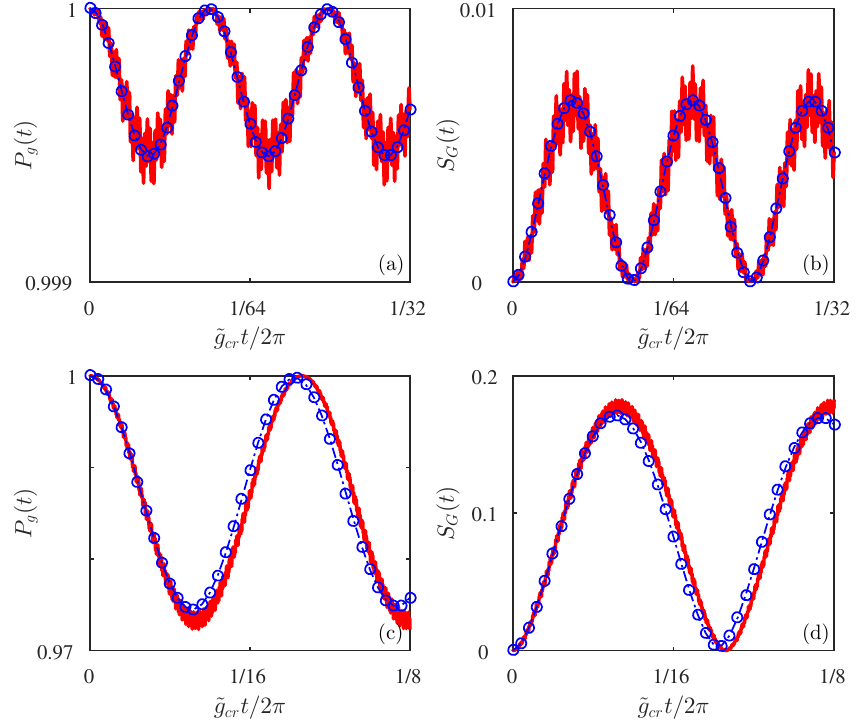}
\end{center}
\caption{The evolution of the atomic ground-state probability $P_g(t)$ (a,c) and the entanglement entropy $S_G(t)$ (b,d) as a function of time, obtained by numerically integrating the original Hamiltonian in \eq{\ref{eq_ht}} (red solid line), and the effective Hamiltonian in \eq{\ref{eq_heff}} (blue dashed lines with circles), respectively. Two sets of parameters are considered: (a, b) $\omega_r = 2\pi \times 15.448$~GHz and $\omega_b = 2\pi \times 20.548$~GHz, $\Omega_r/\hbar=\Lambda_r/\hbar=\Omega_b/\hbar=\Lambda_b/\hbar= 2\pi \times 10.5$~MHz; (c, d) $\omega_r = 2\pi \times 15.0897$~GHz and $\omega_b = 2\pi \times 20.9097$~GHz, $\Omega_r/\hbar=\Lambda_r/\hbar=\Omega_b/\hbar=\Lambda_b/\hbar= 2\pi \times 15$~MHz. This leads to simulated effective parameters of $\tilde{\eta}=300$, $\tilde{g}_{cr}/\tilde{g}_c =\tilde{g}_{r}/\tilde{g}_c = 0.40$ for (a, b); and $\tilde{\eta}=300$, $\tilde{g}_{cr}/\tilde{g}_c =\tilde{g}_{r}/\tilde{g}_c \approx 2.88$ for (c, d). For the simulation, the system is initially prepared in the ground state of the whole system $\ket{g,0}$, and the rest of the parameters are chosen as $\omega = 2\pi \times 3$~GHz, $\omega_q = 2\pi \times 18$~GHz, and $g = 2\pi \times 37$~MHz~\cite{liu_superconducting_2007}.}
\label{fig_psqpt}
\end{figure}
In the following, with realistic parameters in circuit QED systems \cite{liu_superconducting_2007,huang_optimal_2014}, we demonstrate that our proposal is capable of simulating the AQRM with finite large $\tilde{\eta}$ for studying quantum phase transitions, i.e. $\tilde{\eta} = 300$ is achieved in \fig{\ref{fig_psqpt}}. As an illustration, we make comparisons between the original Hamiltonian in \eq{\ref{eq_ht}} and the effective Hamiltonian in \eq{\ref{eq_heff}} for the ground-state probability $P_g(t) =\abs{\langle{\Psi_G}\vert{g}\rangle}^2$ and the ground-state entanglement entropy $S_G=-\Tr{\big[\rho_G^q \log_2(\rho_G^q )\big]}$, where $\ket{\Psi_G}$ is the ground state of the total system, $\rho_G^q = \Tr_f{\big[\rho_G\big]}$ is the reduced density matrix of the qubit's subsystem by tracing out the field's degree of freedom. The ground-state probability $P_g(t)$ indicates the atomic-excitation probability in the ground state $\ket{\Psi_G}$ of the total system, and the entanglement entropy $S_G$ measures the entanglement between the qubit and the resonator.


Clearly shown in \fig{\ref{fig_psqpt}}, the results from the original Hamiltonian (red solid lines) are completely consistent with the ones from the effective Hamiltonian (blue dashed lines with circles) for the two sets of parameters: (a, b) $\omega_r = 2\pi \times 15.448$~GHz and $\omega_b = 2\pi \times 20.548$~GHz, $\Omega_r/\hbar=\Lambda_r/\hbar=\Omega_b/\hbar=\Lambda_b/\hbar= 2\pi \times 10.5$~MHz; (c, d) $\omega_r = 2\pi \times 15.0897$~GHz and $\omega_b = 2\pi \times 20.9097$~GHz, $\Omega_r/\hbar=\Lambda_r/\hbar=\Omega_b/\hbar=\Lambda_b/\hbar= 2\pi \times 15$~MHz, respectively. It is also obvious that such strong driving amplitudes of a few megahertz are sufficient enough to simulate the dynamics USC and even beyond. Comparison of \fig{\ref{fig_psqpt}}(b, d) indicates that the ground state $\ket{\Psi_G}$ remains in the product state $\ket{g,0}$ for coupling strength $\tilde{g}_{r}=\tilde{g}_{cr}$ smaller than the critical coupling $\tilde{g}_{c}$, i. e., $\tilde{g}_{cr}/\tilde{g}_{c}=0.4$ in \fig{\ref{fig_psqpt}}(b), and the ground state $\ket{\Psi_G}$ evolves to an entangled state for coupling strength $\tilde{g}_{r}=\tilde{g}_{cr}$ larger than the critical coupling $\tilde{g}_{c}$, i. e., $\tilde{g}_{cr}/\tilde{g}_{c}=2.88$ in \fig{\ref{fig_psqpt}}(d). This agrees with the results in \fig{\ref{fig_fs}}, where the phase transitions appears at the critical point of $\tilde{g}_c = 1$~\cite{hwang_quantum_2015,wei2018prab,liu_universal_2017}.

\section{Quantum information with degenerate AQRM}
\label{sec4}
Without lose of generality, our scheme of simulating the controllable AQRM can be generalized to the multi-qubit case, where multiple flux qubits are coupled to the LC circuit as shown in \fig{\ref{fig_setup}}(b).
When the corresponding conditions for each qubit to realize the effective anisotricpic Rabi model as in \eq{\ref{eq_heff}} are well satisfied, and the parameters for the $l$th qubit $(l=1,2,...N)$ are chosen such that, $\delta_b^l = -\delta_r^l = \delta$,  $\varphi_b^l = -\varphi_r^l = \varphi$, and $\Lambda_r^l = \Lambda_b^l =\Lambda$, the effective multi-qubit Hamiltonian can be written as
\beq
{\mathcal H}_{\delta,\varphi}^N = \bar{g} J_x \left( a \, e^{-i \delta t -i \varphi } + a^\dag\, e^{i \delta t +i \varphi}\right),
\label{eq_nqmHd0phi}
\eeq
where $J_x = \sum_{l=1}^N \sigma_x^l$ are the collective qubit operators and we have defined $\bar{g} \equiv \Lambda/2$.
Hereafter we name the case of degenerate qubits with $\tilde{\om}_q^l=0$ as the degenerate AQRM.

The evolution operator ${\mathcal U}_{\delta,\varphi}$ can then be found as
\beq
{\mathcal U}_{\delta,\varphi} = \exp{[i \bar{\Phi} (t)\,J_x^2]} \, {\mathcal D} \Big(\alpha (t) J_x \Big),
\label{eq_uoperator}
\eeq
where $\bar{\Phi} (t)= {(\bar{g} /\delta)}^2 (\delta t - \sin{\delta t})$
and ${\mathcal D} (\chi) = \exp{[\chi(t) a^\dag - \chi^*(t)a]}$ is the displacement operator with $\chi(t)$ being the collective displacement amplitude of the oscillator.

\subsection{The generation of macroscopic Schr\"{o}dinger cat states}
\begin{figure}[!b]
\begin{center}
\includegraphics[scale=1]{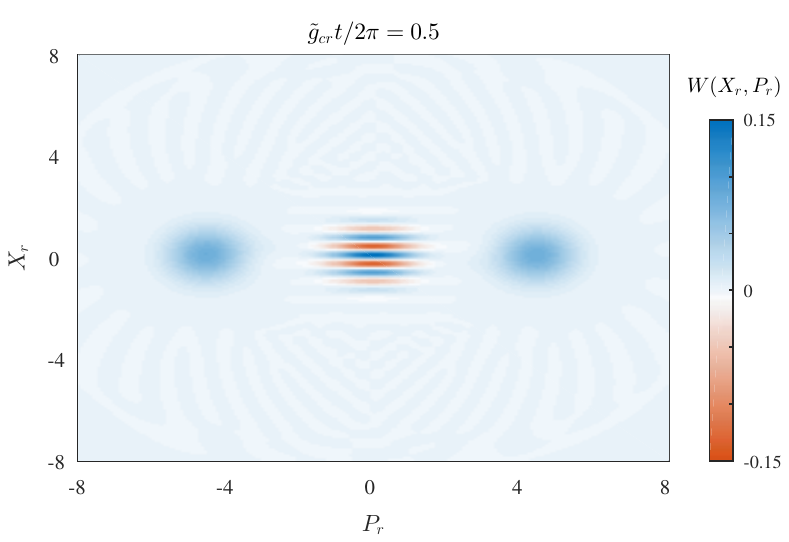}
\end{center}
\caption{The Wigner function of the macroscopic Schr\"{o}dinger cat state generated from a projective measurement on the qubit state $\ket{g}$ in the case of $\tilde{\omega}_q=\tilde{\omega}=0$, $\tilde{g}_r =1$, $\tilde{g}_{cr}=1$, which is calculated {\it ab initio} from \eq{\ref{eq_ht}} with $\ket{g,0}$ being the initial state of the system. The parameters for this plot are chosen as $\omega = 2\pi \times 3$~GHz, $\omega_q = 2\pi \times 18$~GHz, $\omega_r = 2\pi \times 15$~GHz and $\omega_b = 2\pi \times 21$~GHz, $\Omega_r/\hbar=\Lambda_r/\hbar=\Omega_b/\hbar=\Lambda_b/\hbar= 2\pi \times 15$~MHz, and $g = 2\pi \times 37$~MHz~\cite{liu_superconducting_2007}.}
\label{fig_cat}
\end{figure}
It has been proved that, the Schr\"{o}dinger cat states have promising applications in hardware-efficient
quantum memory and quantum error corrections~\cite{leghtas_hardware-efficient_2013,ofek_extending_2016}. In this following, we show the performance of our scheme in generating this class of non-classical states with both theoretical and numerical approaches. In the single-qubit case, $J_x$ in \eqs{\ref{eq_nqmHd0phi},\ref{eq_uoperator}} are replaced by $J_x=\sigma_x$, and when the initial state of the whole system is prepared in the ground state as $\ket{\Psi_1(0)}=\ket{g,0}$, we obtain the final state at time $t$ as
\beq
\ket{\Psi_1(t)} =e^{i \bar{\Phi} (t)}\frac{1}{\sqrt{2}} \big(\ket{+}\ket{\alpha} - \ket{-}\ket{-\alpha}\big),
\label{eq_1qcoh}
\eeq
where $\ket{\pm}=\left(\ket{e} \pm \ket{g}\right)/\sqrt{2}$, and $\ket{\pm\alpha(t)}$ being the coherent states of the harmonic oscillator, which are of the same amplitude but opposite phase in the phase space. $\alpha(t) = {\bar{g}} (1-e^{i\delta t}) e^{i\varphi}/{\delta}$ is the coherent-state amplitude for the single-qubit case.
It is worthy noting that from \eq{\ref{eq_1qcoh}} that, since $\sigma_x^2=I$, the first term in \eq{\ref{eq_uoperator}} behaves only as a global phase factor in \eq{\ref{eq_1qcoh}}. Obviously, depending on the states of the flux qubit $\ket{\pm}$, the coherent states undergo different displacements $\ket{\pm\alpha(t)}$, respectively. In the bases of the $\{\ket{e},\,\ket{g}\}$, the state in
\eq{\ref{eq_1qcoh}} can be rewritten as
\beq
 \ket{\Psi_1(t)} =\frac{1}{2\mathcal N}\left(\ket{e} \ket{{\mathcal C}^-_\alpha} + \ket{g} \ket{{\mathcal C}^+_\alpha} \right),
\label{eq_1qcat1}
\eeq
where $\ket{{\mathcal C}^\pm_\alpha } \equiv {\mathcal N}(\ket{\alpha} \pm \ket{-\alpha})$ with ${\mathcal N} = 1/\sqrt{2 (1\pm e^{-2\abs{\alpha}^2})} \approx 1/\sqrt{2}$ are the so-called even $(\ket{{\mathcal C}^+_\alpha})$ and odd $(\ket{{\mathcal C}^-_\alpha } )$ Schr\"{o}dinger cat states. By choosing the phase $\varphi$, the four types of the quasi-orthogonal states $\ket{\pm \alpha}$ and $\ket{\pm i \alpha}$ \cite{leghtas_hardware-efficient_2013}, i.e., $\abs{\langle{\alpha}|{i \alpha}\rangle}^2<<1$ (note that for $\alpha=2$, $\abs{\langle{\alpha}|{i \alpha}\rangle}^2<10^{-3}$), can be generated by measuring the qubit in the $\ket{\pm}$ bases.
By performing projective measurements in the qubit $\{\ket{e},\ket{g}\}$ bases, the oscillator will collapse into the Schr\"{o}dinger cat states with probability of ${(1\pm e^{-2\abs{\alpha}^2})}/2$, respectively. As shown in \eq{\ref{eq_1qcat1}}, the even cat state $\ket{{\mathcal C}^+_\alpha}$ is generated with a projective measurement onto the qubit's ground state $\ket{g}$. Seen from \eq{\ref{eq_1qcat1}} that, the maximum displacement amplitude is $\abs{\alpha}_{max}=2\bar{g} /\delta$, and it can be obtained at the times $t = (2m+ 1) \pi/\delta$ for natural number $m$. By choosing a small value for $\delta$ and a large effective coupling strength $\bar{g}$, we can create macroscopically distinct Schr\"{o}dinger cat states of considerable size of $\abs{\alpha}>1$.

In another aspect, the displacement amplitude of the Schr\"{o}dinger cat states can be further enhanced with even number of flux qubits by exploring the multi-qubit case and preparing the system in the state of $\ket{\Psi_N(0)}=(\ket{+,+,...,+}-\ket{-,-,...,-}) \ket{0}/\sqrt{2}$. In this case, the state after evolution is given by
\beq
\ket{\Psi_N(t)}=\frac{e^{i N^2 \bar{\Phi} (t)}}{\sqrt{2}} \big(\ket{+,+,...,+} \ket{N\alpha} - \ket{-,-,...,-}\ket{-N\alpha}\big),
\label{eq_nqcoh}
\eeq
where the coherent-state amplitude is enhanced by a factor $N$, and the first term in \eq{\ref{eq_uoperator}} remains as a global phase factor. However, collective measurements on the flux qubit in the bases of $(\ket{+,+,...,+}\pm\ket{-,-,...,-})/\sqrt{2}$ are required to obtain the Schr\"{o}dinger cat states with an enhanced amplitude.

\subsection{The two-qubit controlled quantum phase gate generation}
As seen from the evolution operator \eq{\ref{eq_uoperator}}, the Hamiltonian in \eq{\ref{eq_nqmHd0phi}} introduces qubit-qubit interaction between any pair of qubits. And thus our circuit can be used to generate quantum gates and produce highly-entangled states between qubits. Let the system evolve for a time period of $T = 2 \pi/\delta$, we obtain $\chi(T)=0$ and up to an overall phase factor, the evolution operator can be recast as 
\beq
U(T)=\exp{i\left[\bar{\theta}\sum_{k>l=1}^N \sigma_x^l \sigma_x^k\right]},
\label{eq_2quoperator}
\eeq
with $\bar{\theta}=2 \bar{\Phi} (T)=4\pi \bar{g}^2/\delta^2$. In the following, we show that the generation of a two-qubit quantum controlled-NOT gate is straightforward from \eq{\ref{eq_2quoperator}}. In the two-qubit bases of $\{\ket{ee},\ket{eg},\ket{ge},\ket{gg}\}$, the evolution operator can be expressed as
\beq
U(T)=\left(
\begin{array}{cccc}
 \cos\bar{\theta} & 0 & 0 & i \sin\bar{\theta}\\
 0 & \cos\bar{\theta}& i \sin\bar{\theta}& 0 \\
 0 & i \sin\bar{\theta} & \cos\bar{\theta} & 0 \\
 i \sin\bar{\theta} & 0 & 0 & \cos\bar{\theta} \\
\end{array}
\right),
\eeq
which represents the non-trivial two-qubit gates when $\bar{\theta}\neq m \pi$θ $(m = 0, ±1, ±2, \cdots)$. Specifically, when $\bar{\theta} = \pi/4$ (i.e., $\bar{g}=\delta/4)$, $U(T)$ is locally equivalent to the controlled-NOT (CNOT) gate.

\subsection{The simulated degenerate AQRM}
\begin{figure}[!t]
\begin{center}
\includegraphics[scale=1.2]{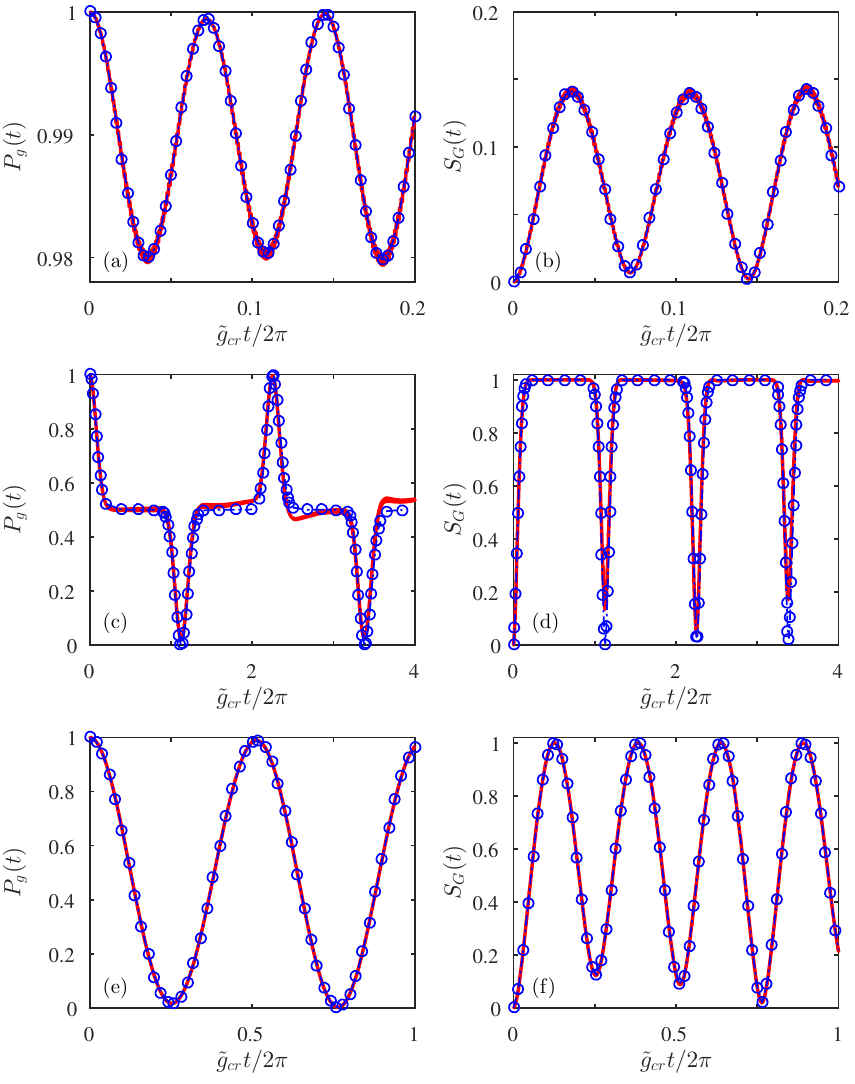}
\end{center}
\caption{The evolution of the ground-state probability $P_g(t)$ (a,c,e) and the entanglement entropy $S_G(t)$ (b,d,f) as a function of time, obtained by numerically integrating the original Hamiltonian in \eq{\ref{eq_ht}} (red solid line), and the effective Hamiltonian in \eq{\ref{eq_heff}} (blue dashed lines with circles), respectively. We have considered three cases: (a, b) $\Omega_r/\hbar=\Lambda_r/\hbar= 2\pi \times 15$~MHz, $\Omega_b/\hbar=\Lambda_b/\hbar= 2\pi \times 3$~MHz; (c, d) $\Omega_r/\hbar=\Lambda_r/\hbar= 2\pi \times 15$~MHz, $\Omega_b/\hbar=\Lambda_b/\hbar= 2\pi \times 15$~MHz; (e, f) $\Omega_r/\hbar=\Lambda_r/\hbar= 2\pi \times 3$~MHz, $\Omega_b/\hbar=\Lambda_b/\hbar= 2\pi \times 15$~MHz. This lead to simulated effective parameters of $\tilde{\omega}_q=\tilde{\omega}=0$ with $\tilde{g}_r=1$, $\tilde{g}_{cr} =0.2$ for (a, b); $\tilde{g}_r=\tilde{g}_{cr}=1$ for (c, d); and $\tilde{g}_{r} =0.2$, $\tilde{g}_{cr}=1$ for (e, f). For this simulation, the system is also initially prepared in the ground state of the whole system $\ket{g,0}$, and the rest of the parameters are chosen to the same as for \fig{\ref{fig_cat}}.}
\label{fig_psw0wa0_3by2}
\end{figure}
\begin{figure}[!b]
\begin{center}
\includegraphics[scale=1.2]{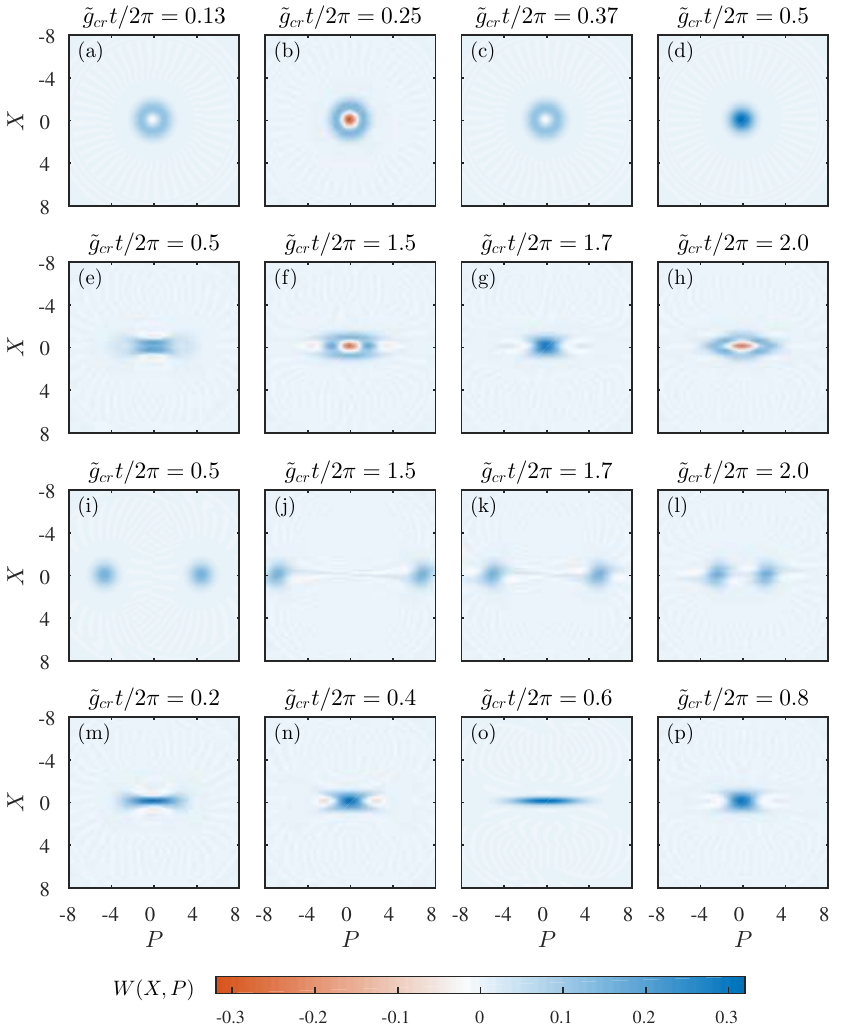}
\end{center}
\caption{The Wigner function $W(X,P)$ of the field state at different interaction times after tracing out the qubit's degree of freedom, calculated {\it ab initio} from \eq{\ref{eq_ht}} with $\ket{g,0}$ being the initial state of the system. We have considered four cases: (a-d) $\Omega_r=\Lambda_r= 0$ and $\Omega_b/\hbar=\Lambda_b/\hbar= 2\pi \times 15$ MHz, which corresponds to simulated effective parameters of $\tilde{\omega}_q=\tilde{\omega}=\tilde{g}_r =0$, $\tilde{g}_{cr}=1$; (e-h) $\Omega_r/\hbar=\Lambda_r/\hbar= 2\pi \times 7.5$~MHz and $\Omega_b/\hbar =\Lambda_b/\hbar = 2\pi \times 15$~MHz, which corresponds to simulated effective parameters of $\tilde{\omega}_q=\tilde{\omega}=0$, $\tilde{g}_r =0.5$, $\tilde{g}_{cr}=1$; (i-l)  $\Omega_r=\Lambda_r= 2\pi \times 15$~MHz and $\Omega_b/\hbar =\Lambda_b/\hbar= 2\pi \times 15$~MHz, which corresponds to simulated effective parameters of $\tilde{\omega}_q=\tilde{\omega}=0$, $\tilde{g}_r =\tilde{g}_{cr}=1$;  (m-p) $\Omega_r/\hbar=\Lambda_r/\hbar=2\pi \times 15$~MHz and $\Omega_b/\hbar =\Lambda_b/\hbar= 2\pi \times 7.5$~MHz, which corresponds to simulated effective parameters of $\tilde{\omega}_q=\tilde{\omega}=0$, $\tilde{g}_r =1$, $\tilde{g}_{cr}=0.5$. For this simulation, the system is also initially prepared in the ground state of the whole system $\ket{g,0}$, and the rest of the parameters are chosen to the same as for \fig{\ref{fig_cat}}.}
\label{fig_wigner}
\end{figure}
By the numerical calculations with practical parameters~\cite{liu_superconducting_2007}, we prove that our proposal serves well in simulating the degenerate AQRM. Without loss of generality, we display in \fig{\ref{fig_psw0wa0_3by2}} the simulation of the double degenerate AQRM, where both $\tilde{\omega}_q$ and $\tilde{\omega}$ are zeros. The atomic-ground-state probability $P_g(t)$ and the ground-state entanglement entropy $S_G$ are plotted for three sets of parameters, $\Omega_r/\hbar=\Lambda_r/\hbar= 2\pi \times 15$~MHz, $\Omega_b/\hbar=\Lambda_b/\hbar= 2\pi \times 3$~MHz for \fig{\ref{fig_psw0wa0_3by2}}(a, b); $\Omega_r/\hbar=\Lambda_r/\hbar= 2\pi \times 15$~MHz, $\Omega_b/\hbar=\Lambda_b/\hbar= 2\pi \times 15$~MHz for \fig{\ref{fig_psw0wa0_3by2}}(c, d); and $\Omega_r/\hbar=\Lambda_r/\hbar= 2\pi \times 3$~MHz, $\Omega_b/\hbar=\Lambda_b/\hbar= 2\pi \times 15$~MHz for \fig{\ref{fig_psw0wa0_3by2}}(e, f). The frequency of the red and blue drivings are chosen to be $\omega_r = 2\pi \times 15$~GHz and $\omega_b = 2\pi \times 21$~GHz. The curved lines for the original Hamiltonian in \eq{\ref{eq_ht}} (red solid line) reproduce the ones calculated for the effective Hamiltonian in \eq{\ref{eq_heff}} (blue dashed lines with circles) with high accuracy. The numerical agreements shown in both \fig{\ref{fig_psqpt}} and \fig{\ref{fig_psw0wa0_3by2}} prove that our scheme has excellent performance in simulating static properties and the dynamics of the double AQRM in both the USC and the DSC regimes.


What coming along with the atomic population transfers are the collapses and revivals of the photon wave packets and the variation of the photon statistics. In the following, by employing the Wigner quasi-probability distribution function (WF), we show some interesting features of the field statistical properties of  the double degenerate AQRM with $\tilde{\om}=\tilde{\om}_q=0$. 
In \fig{\ref{fig_wigner}}, we plot the WF of the AQRM at different time intervals for four sets of parameters with the initial state $\ket{g,0}$ and $\omega_r = 2\pi \times 15$~GHz and $\omega_b = 2\pi \times 21$~GHz. The top row of \fig{\ref{fig_wigner}}(a-d) depicts the evolution of the WF of the field generated when $\Omega_r=\Lambda_r= 0$, $\Omega_b/\hbar=\Lambda_b/\hbar= 2\pi \times 15$ MHz, which corresponds to the population transfer between the states of $\ket{g,0}$ and $\ket{e,1}$, and the WF of the single photon Fock state is shown in \fig{\ref{fig_wigner}}(c) at time $\tilde{g}_{cr} t/{2\pi}=0.25$.
The third row of \fig{\ref{fig_wigner}}(i-l) shows the evolution of the WF of the field generated when $\Omega_r/\hbar=\Lambda_r/\hbar=\Omega_b/\hbar=\Lambda_b/\hbar= 2\pi \times 15$~MHz, which describes a mixture of two coherent states $\ket{\pm \alpha}$ with time-dependent displacement amplitude of $\alpha (t) = -i \bar{g}_r t$~\cite{ashhab_qubit-oscillator_2010}. The amplitudes of the coherent states ideally increase linearly and practically, they will be prevented from diverging into instability by the damping of the oscillator and the finite duration of the evolution. It is noted that the small distortion of the WF from the ones of the ideal coherent state is due to a small deviation of our scheme from the effective ones for longer evolution time.
The second row and the bottom row of \fig{\ref{fig_wigner}} display the field properties with unbalanced and nonzero rotating and counter-rotating coupling terms in the degenerate AQRM, i. e., $\Omega_r/\hbar=\Lambda_r/\hbar= 2\pi \times 7.5$~MHz and $\Omega_b/\hbar =\Lambda_b/\hbar = 2\pi \times 15$~MHz for \fig{\ref{fig_wigner}}(e-h), and $\Omega_r/\hbar=\Lambda_r/\hbar=2\pi \times 15$~MHz, $\Omega_b/\hbar =\Lambda_b/\hbar= 2\pi \times 7.5$~MHz for \fig{\ref{fig_wigner}}(m-p). In these two cases, both the rotating and counter-rotating terms contribute to the dynamics of the system, but unbalanced. An intuitive picture to understand these figures could be the following. Started from $\ket{g,0}$, the photons spread independently along the even parity chain, and thus produce a qubit-resonator entangled state. Such entangled state has the properties of the displaced squeezed states, whose squeezing parameters are functions of the relative ratio $\tilde{g}_{cr}/\tilde{g}_r$.

\section{Conclusions}
\label{sec5}

In summary, by manipulating the flux qubits with bichromatic time-dependent magnetic fields, we propose an experimentally-accessible method to approach the physics of the anisotropic quantum Rabi model (AQRM) in broad parameter ranges. 
With all-round tunability of the AQRM, we investigate its rich applications for quantum phase transitions (QPTs) from the perspective of information metric. Universal information like critical exponents can be well extracted from the scaling behavior of the fidelity susceptibility. Despite the differences in the anisotropy and the frequency size, a fixed point of a cumulant ratio is predicted at the critical point of the QPTs and a universal scaling of the cumulant ratio is obtained with appropriate rescaling of the parameters. With numerical calculations we prove that our proposal is capable of achieving the parameter ranges of demonstrating the quantum phase transition with finite large frequency scale. Moreover, we find that our scheme severs well for the generation of the macroscopic Schr\"{o}dinger cat states and the quantum controlled phase gates. Hence, our scheme serves as a favorable platform to explore the Rabi physics and testify the universal scaling of quantum critical phenomena in few-body systems. Especially, our scheme may also open the appealing possibility of
experimentally exhibiting the gapless Nambu-Goldstone mode, which appears in the pure Jaynes-Cummings model with sufficiently strong coupling strength. This is forbidden in natural systems with very large coupling due to the failure of the rotating-wave approximation.

\vspace{8pt}
\section{Acknowledgements}
This work was supported by the NSFC under Grant Nos.~11404407, 11474211, 11325417, 11674139, 11604009, the Jiangsu NSF under Grant Nos.~BK20140072,~BK20141190, and the China Postdoctoral Science Foundation under Grant Nos.~2015M580965 and 2016T90028. G.R. acknowledges the support from FONDECYT under grant No.~1150653.

\section*{References}
\bibliographystyle{iopart-num}
\bibliographystyle{iop}
\bibliographystyle{iopart}
\bibliography{aniso_rabi_NJP}

\end{document}